\documentstyle[pra,aps]{revtex}
\begin{document}
\draft

\twocolumn[\hsize\textwidth\columnwidth\hsize\csname 
@twocolumnfalse\endcsname

\title{Zeta function method and repulsive Casimir forces for an unusual pair
of plates  at finite temperature}

\author{F. C. Santos, A. Ten\'orio and A. C. Tort\cite{email}}

\address{Instituto de F\'{\i}sica, Universidade Federal do Rio de Janeiro,
Caixa Postal 68528, 21945-970 Rio de Janeiro RJ, Brazil}

\date{\today}
\maketitle

\begin{abstract}
We apply the generalized Zeta function method to compute the Casimir energy
and pressure between an unusual pair of parallel plates  at finite
temperature, namely, a perfectely conducting plate
($\epsilon\rightarrow\infty$) and an infinitely permeable one
($\mu\rightarrow\infty$). The high and low temperature limits of these
quantities are discussed.  Relationships between high and low temperature
limits for the free energy are established by means of a modified version of
the temperature inversion  symmetry.
\end{abstract}

\vskip2pc] \label{sec:level1}

\section{Introduction}
Since Casimir's paper \cite{Ca} on the attraction between two parallel
perfectly conducting plates due to the vacuum fluctuations of the
electromagnetic field was published, a considerable amount of effort, which
varies from the investigation of new geometries and theories to the
application of the Casimir effect to alternative technologies, has been put
into the study of this importante subject.  (For a review see, for example,
Mostepanenko and Trunov \cite{MosteTrunov} or Plunien {\it et al}).
Recently, the experimental observation of this effect was greatly improved
by the experiments due to Lamoreaux and Mohideen and Roy \cite{Lamo}. From
the theoretical viewpoint Casimir's approach to this problem essentially
consisted in computing the interaction energy between the plates as the
regularized difference between the zero point energies with and without
boundary conditions dictated by the physical situation at hand, for
instance, the perfectly conducting character of the plates. The great
novelty of Casimir's 1948 paper was not the fact that two neutral object
were attracted towards each other, which was familiar to those studying
dispersive van der Waals forces, but the simplicity of the method of
calculating this attraction in the framework of quantum field theory.

 Casimir's definition of the vacuum energy requires a regularization recipe
for its implementation. Many regularization techniques are available
nowadays and, depending on the specific physical situation at hand, one of
them may be more suitable than the others. Particularly, methods of
computing effective actions are in general very powerful to give physical
meaning to the divergent quantities we must deal with. Here we will be
concerned with one of these methods, namely, the so-called generalized zeta
function method \cite{ZETA}. There are several examples of the application
of this method to the evaluation of the Casimir effect at zero and finite
temperature, in its global and also in its local version, see for example,
\cite{Actor}. Here we will apply it to the case of a pair of parallel
infinite plates one of which is perfectly conducting ($\epsilon\to\infty$),
while the other is infinitely permeable ($\mu\to\infty$). The setup will be
considered to be in thermal equilibrium with a heat reservoir at finite
temperature $T$. This problem at zero temperature was analyzed by Boyer
\cite{Boyer} two decades ago in the framework of random electrodynamics, a
kind of classical electrodynamics which includes the zero-point
electromagnetic radiation. Boyer was able to show that for this unusual pair
of plates the Casimir energy is positive which results in a repulsive force
per unit area between the plates. Recently Boyer's result at zero
temperature was rederived by zeta function methods \cite{CP}. Repulsive
electromagnetic Casimir forces can arise in geometrically more complicated
setups as for instance a spherical shell. More complicated geometries may
lead to extremely involved calculations. Though for a given geometry it is
possible to infer on dimensional grounds only, the form of the Casimir
energy, its correct algebraic sign, that is: the attractive or repulsive
character of the associated Casimir force, and numerical factors are
obtained only after complex calculations.  More generally, the non-trivial
dependence of the algebraic sign of the Casimir force on the type of quantum
field being studied, type of spacetime, on the dimensions of the spacetime
and on the type of boundary imposed on the quantum field  was denoted by
some authors as `the mystery of the Casimir force' \cite{E&R91}.

Boyer's unsual pair of plates is the simplest example where we can find
repulsive Casimir forces at work. They were recently employed by Hushwater
\cite{Hushwater} as a counterexample in order to show that the naive
interpretation of the standard Casimir attraction between two parallel
conducting plates as being due to a difference between the number of vacuum
modes in the region between the plates and the region outside the plates
does not apply. Boyer's plates were also used in connection with the
Scharnhorst effect \cite{Scharnhorst} where they provided one more example
in which the propagation of a light signal in the confined electromagnetic
vacuum is modified with respect to propagation in the unconstrained vacuum
\cite{CPFST}. While thermal corrections to the standard Casimir effect were
calculated by several authors and constitute a large body of literature on
this subject, see \cite{THERMALCASIMIR} for representative examples, there
are to our knowledge no calculations of thermal corrections to the repulsive
Casimir effect associated with Boyer's  setup. Here, in order to remedy this
situation we take into account the thermal effects of the equilibrium state,
and study this problem within the framework of finite temperature QFT.
Moreover, we show that, though boundary conditions are not symmetric, it is
still possible to discuss temperature inversion symmetry for this system. We
take advantage of the fact that in the case of the simple geometry we are
considering the electromagnetic field can be simulated by a scalar massless
field. The insertion of a multipicative factor equal to 2 will take into
account the two possible polarizations of the electromagnetic field. The
article is divided as follows: Firstly we derive general expressions for the
free energy and pressure. Secondly, we consider the low and high
temperatures limits of these quantities. Thirdly, we show that Boyer's setup
is equivalent to the difference beteween two Casimir's setups --- a fact
that allows us to discuss the temperature inversion symmetry associated with
this system. Finally, the last section is devoted to concluding remarks. We
will employ units such that Boltzmann constant, the speed of light and
$\hbar=h/2\pi$ are set equal to unity.
\section{Evaluation of the free energy}
Since we will be dealing with a system in thermal equilibrium, the imaginary
time formalism will be convenient. In order to apply the generalized zeta
function method, let us introduce the partition function ${\cal Z}$ for a
bosonic theory \cite{Kapusta}:
\begin{equation}
{\cal Z} = N\int_{\mbox{\tiny Periodic}}
[D\phi]\exp{\left(\int_0^\beta\,d\tau\int d^3x\,{\cal L}\right)}\,,
\end{equation}
where ${\cal L}$ is the Lagrangian density for the theory under
consideration, $N$ is a constant and `periodic' means that the functional
integral is to be evaluated over field configurations satisfying:
\begin{equation}\label{PC}
\phi(x,y,z,0)=\phi(x,y,z,\beta)\,, 
\end{equation}
where $\beta=T^{-1}$, the reciprocal of the temperature, is the periodic
length in the Euclidean time axis. The Helmholtz free energy $F(\beta)$ is
related to the partition function ${\cal Z}(\beta)$ through the relation
$F(\beta)=-\beta^{-1}\log{{\cal Z}(\beta)}$. Other than the periodic
conditions given by (\ref{PC}), we must also consider boundary conditions
which are determined by the geometry and the nature of the physical system
under study. An example is the configuration mentioned  above. Choosing
Cartesian axes such that the axis $OZ$ is perpendicular to both plates with
the perfectly conducting plate at $z=0$ and the infinitely permeable one at
$z=d$, the boundary conditions on the vacuum oscillations of the
electromagnetic field are the following: the tangential components of the
electric field as well as the normal component of magnetic field must vanish
at $z=0$, while the tangential components of the magnetic field and the
normal component of the electric field must vanish at $z=d$. As mentioned
before, for the plate geometry that we are considering, the electromagnetic
field can be mimicked by a scalar massless field $\phi$. The boundary
conditions stated above can be translated into:

\begin{equation}\label{MC}
\phi(\tau, x,y,z=0)=0\,;\;\;\;\;\;{\partial\phi(\tau,x,y,z=d)\over\partial
z}=0\,,
\end{equation}
where $\tau$ is the Euclidean time. The insertion at the end of the
calculation of a factor $2$ will take into account the two possible
transverse polarizations of the electromagnetic field. Thus we write
$\log{{\cal Z}(\beta)}$ as:
\begin{equation}\label{PFU}
\log{{\cal Z}(\beta)}=\left(-{1\over
2}\right)\log\det\left(-\partial_{\mbox{\tiny E}}|{\cal F}_d\right)\,,
\end{equation}
where $\partial_{\mbox{\tiny E}}=\partial^2/\partial\tau^2+\nabla^2$, and
the symbol ${\cal F}_d$ stands for the set of functions which satisfy
conditions (\ref{PC}) and (\ref{MC}). The generalized zeta function method
basically consists of the following three steps: (i) first, we compute the
eigenvalues of the operator $-\partial_{\mbox{\tiny E}}$ whose
eigenfunctions obey the appropriate boundary conditions and write $\zeta
(s;-\partial_{\mbox{\tiny E}})=\mbox{Tr}\,(-\partial_{\mbox{\tiny
E}})^{-s}$; (ii) second, we perform an analytical continuation of
$\zeta(s;-\partial_{\mbox{\tiny E}})$ to a meromorphic function on the whole
complex $s$-plane; (iii) finally, we compute $\det{(-\partial_{\mbox{\tiny
E}}|{\cal F}_d)}=\exp{\left(-{\partial\zeta (s=0;-\partial_{\mbox{\tiny
E}})\over\partial s}\right)}$. Combining equation (\ref{PFU}) with the
definition of free energy we obtain:
\begin{equation}
F(\beta)=-\beta^{-1}{\partial\zeta (s=0;-\partial_{\mbox{\tiny
E}})\over\partial s}\,.
\end{equation}
The eigenvalues of $-\partial_{\mbox{\tiny E}}$ whose eigenfunctions satisfy
(\ref{PC}) and (\ref{MC}) are:
\begin{equation}
\left\{k_x^2+k_y^2+\left(n+{1\over 2}\right)^2{\pi^2\over
d^2}+{4\pi^2m^2\over \beta^2},\right\},
\end{equation}
where $k_x,k_y \in\Re$, $ n\in \left\{0,1,2,3,...\right\}$ and $m\in\left\{
0,\pm 1, \pm 2,...\right\}$. The generalized zeta function then reads:
\begin{eqnarray}
\zeta (s,-\partial_{\mbox{\tiny E}})& = &
L^2\sum_{m=-\infty}^\infty\sum_{n=0}^\infty\int{dk_xdk_y\over
(2\pi)^2}\times \nonumber\\
&\times&\left[k_x^2+k_y^2+(2n+1)^2{\pi^2\over
4d^2}+{4\pi^2m^2\over\beta^2}\right]^{-s}\;.
\end{eqnarray}
where $L^2$ is the area of the plates. After rearranging terms in the
summations, changing to polar coordinates and integrating the angular part
out, we can rewrite this last equation as:
\begin{eqnarray}
& \zeta & (s,-\partial_{\mbox{\tiny E}}) = {L^2\over
2\pi}\left\{\sum_{n=1}^{\infty\;\;\prime}\int_0^\infty
dk\,k\left[k^2+{n^2\pi^2\over 4d^2}\right]^{-s}\right. \nonumber \\
& + & \left. 2\sum_{n=1}^{\infty\;\;\prime}\sum_{m=1}^\infty
dk\,k\left[k^2+{n^2\pi^2\over
4d^2}+{4\pi^2m^2\over\beta^2}\right]^{-s}\right\}\;,
\end{eqnarray}
where $k^2=k_x^2+k_y^2$ and the prime on the summation symbol remind us that
the integer $n$ assumes odd values only. Using the following representation
for the Euler beta function, {\it c.f.} formula {\bf 3.251}.2 \cite{Grad}:
\begin{eqnarray}
\int_0^\infty dx\,x^{\mu-1}\left(x^2+a^2\right)^{\nu-1} &=& \frac{1}{2}{\cal
B}\left({\mu\over2},1-\nu-{\mu\over2}\right)\nonumber \\
& \times & a^{\mu+2\nu-2}\,,
\end{eqnarray}
where
\begin{equation}
{\cal B}(x,y)={\Gamma(x)\Gamma(y)\over\Gamma(x+y)}
\end{equation}
which holds for $\Re\,\left(\nu+{\mu\over 2}\right)< 1$ and $\Re\,\mu>0$, we
obtain:

\begin{eqnarray}
\zeta (s,-\partial_{\mbox{\tiny E}}) & = & {L^2\over 4\pi}{\Gamma (s-1)\over
\Gamma (s)}\left[\left({\pi\over
2d}\right)^{2-2s}\sum_{n=1}^{\infty\;\;\prime} n^{2-2s}  \right. \nonumber \\
 & + & \left.
2\pi^{2-2s}\sum_{m=1}^\infty\sum_{n=1}^{\infty\;\;\prime}\left[{n^2\over
4d^2}+{4m^2\over\beta^2}\right]\right]^{1-s}
\end{eqnarray}
In order to connect the simple sum on the r.h.s. of the above equation to
the Riemann zeta function $\zeta_R$ we write:

\begin{equation}
\sum_{n=1}^{\infty\;\;\prime} n^{2-2s}=(1-2^{2-2s})\zeta_R(2s-2).
\end{equation}
On the other hand, the double sum can be expressed in terms of Epstein
functions which for any positive integer $N$ and $\Re\,z$ large enough are
defined by \cite{Epstein,Kirsten}:

\begin{eqnarray}
& E &_N^{M^2}(z;a_1,a_2,...,a_N):=\sum_{n_1=1}^{\infty}
\sum_{n_2=1}^{\infty}...\sum_{n_N=1}^{\infty}\nonumber\\
&&{1\over
(a_1n_1^2+a_2n_2^2+...+a_Nn_N^2+M^2)^z},
\end{eqnarray} 
where $a_1,...a_N$ and $M^2$$>0$ and writing:
\begin{eqnarray}
& & \sum_{m=1}^\infty \sum _{n=1}^{\infty\;\;\prime}\left[{n^2\over
4d^2}+{4m^2\over\beta^2}\right]^{1-s}=\sum_{m,n=1}^\infty\left[{n^2\over
4d^2}+{4m^2\over\beta^2}\right]^{1-s}\nonumber \\ 
& &\;\;\;\;\;\;\;\;\;\;\;\;\;\;\;\;\;\;\;\;\;\;\;\;\;\;\;\;\;\;\;\;\;
-\sum_{m, n=1}^\infty\left[{(2n)^2\over 4d^2}+{4m^2\over\beta^2}\right]^{1-s},
\end{eqnarray}
we can write:
\begin{eqnarray}
& &\zeta (s,-\partial_{\mbox{\tiny E}}) = {L^2\over 4\pi}{\Gamma (s-1)\over
\Gamma (s)\pi^{2-2s}} \nonumber \\ & &\;\;\;\;\;\;\;\;\;\;\;\;\;\;
\times\left[\left({1\over 2d}\right)^{2-2s}\left(1-2^{2-2s}\right)\zeta_R
(2s-2)\right. \nonumber \\
& + & \left. 2E_2\left(s-1;{1\over 4d^2},
{4\over\beta^2}\right)-2E_2\left(s-1;{1\over d^2},
{4\over\beta^2}\right)\right].
\end{eqnarray}
The Epstein functions can be analytically continued to a meromorphic
function in the complex plane, (see for example \cite{Kirsten}). For $N=2$
and $M^2=0$ the analytic continuation is given by:
\begin{eqnarray}
& & E_2(z;a_1,a_2) = 
-{a_1^{-z}\over 2}\zeta_R (2z)+{1\over 2}\sqrt{\pi\over a_2}
{\Gamma(z-{1\over 2}) \over \Gamma(z)} \nonumber \\
& &\;\;\;\times\; E_1(z-{1\over 2};a_1)
+ {2\pi^z \over \Gamma(z) a_2^{{z\over 2}+{1\over 4}}}\nonumber \\
& &\;\;\;\times\sum_{n,m=1}^\infty 
{m^{z-(1/2)}\over (a_1 n^2)^{(z-(1/2))/2}}
K_{{1\over 2}-z}\left({2\pi m\over \sqrt{a_2}}\sqrt{a_1n^2}\right).
\end{eqnarray}
Here $K_\nu (z)$ is a Macdonald's function\footnote{We use the terminology
employed by N. N. Lebedev in {\it Special Functions and Their Applications},
Dover Publications, New York, 1972. The function $K_\nu (z)$ is also known
as modified Bessel function of the third kind and Bessel function of
imaginary argument.} defined on the complex $z$-plane cut along the negative
real axis, $[-\infty, 0]$. Performing the appropriate substitutions for $z$,
$a_1$ and $a_2$
and taking advantage of the useful fact that the derivative of the function
$G(s)/\Gamma (s)$ at $s=0$ is simply $G(0)$ for a well-behaved $G(s)$ we obtain:

\begin{eqnarray}
&\zeta&^\prime(s,-\partial_{\mbox{\tiny E}}) = -{7\over 8}\times{\pi^2\beta
L^2\over 720 d^3}+ {L^2\sqrt{2}\over\sqrt{\beta}} \sum
_{n,m=1}^\infty\left({md\over n}\right)^{-{3\over2}}\nonumber \\ 
&\times & \left[2^{-{3\over 2}}K_{3/2}\left({\beta\pi nm\over
2d}\right)-K_{3/2}\left({2\beta\pi nm\over 2d}\right)\right]\,.
\end{eqnarray}

It follows that the Helmholtz free energy per unit area for this peculiar
arrangement is given by:
\begin{eqnarray}\label{FR}
{F\over L^2}&=&{7\over 8}\times{\pi^2\over 720 d^3}
-{\sqrt{2}\over\beta^{{3\over 2}}}\sum_{n,m=1}^\infty\left({md\over
n}\right)^{-{3\over2}}\nonumber \\ &\times &\left[2^{-{3\over
2}}K_{3/2}\left({\beta\pi nm\over 2d}\right)-K_{3/2}\left({2\beta\pi nm\over
2d}\right)\right]\,,
\end{eqnarray}
where we have already accounted for the two possible polarization states.
The first term in (\ref{FR}) represents the regularized repulsive Casimir
energy at zero temperature found by Boyer \cite{Boyer}. Notice that this
term is $-7/8$ times the result obtained for the Casimir effect with
Dirichlet boundary conditions at zero temperature. The second term in
(\ref{FR}) is the contribution to the free energy due to thermal effects and
we can recast it into a more manageable form as is shown next.

The Macdonald's functions $K_\nu (z)$ of half-integral order are given by
({\it c.f.} formula {\bf 8.468} in \cite{Grad}):
\begin{equation}\label{BeH}
K_{n+{1\over 2}}(z)=\sqrt{{\pi\over 2z}}e^{-z}\sum_{k=0}^n{(n+k)!\over
k!(n-k)!(2z)^k}\,.
\end{equation}
Hence, defining the dimensionless variable $\xi$ by $\xi:=d/\pi\beta=Td/\pi$
and making use of (\ref{BeH}), we can recast (\ref{FR}) into the form:
\begin{equation}\label{FREE}
{F(\beta)\over L^2}={7\over 8}\times{\pi^2\over 720d^3}-{1\over
\pi\beta^3}f(\xi)\,,
\end{equation}
where $f(\xi)$ is a dimensionless function, also referred to as scaled free
energy,  defined by the double sum:
\begin{eqnarray}\label{DS}
f(\xi)& := &\sum_{n,m=1}^\infty\left[\left({1\over m^3}+{n\over 2\xi
m^2}\right)e^{-nm/2\xi}\right. \nonumber \\
& - & \left.\left({1\over m^3}+{n\over \xi m^2}\right)e^{-(nm/\xi)}\right]\,.
\end{eqnarray}
The sum over $n$ can be readly evaluated and after some manipulations we end
up with:
\begin{eqnarray}\label{FR2}
f(\xi)={1\over 4\xi}\sum_{n=1}^\infty{ \left[{2\xi\over
n}+\coth{\left({n\over 2\xi}\right)}\right]\over n^2\sinh{\left({n\over
2\xi}\right)} }\,.
\end{eqnarray}
Equation (\ref{FR2}) summarizes all thermodynamical information concerning
the bosonic  excitations confined between the plates.
From (\ref{FR2}) we can easily obtain the low temperature regime of the free
energy. It suffices to set $\coth{\left({n\over 2\xi}\right)}\approx 1$,
$\sinh{\left({n\over 2\xi}\right)}\approx 2/\exp{{\left({n\over
2\xi}\right)}}$ and keep the term corresponding to $n=1$:
\begin{equation}
f(\xi\ll 1)\approx \left(1+{1\over 2\xi}\right)\exp{(-{1\over 2\xi})}\,.
\end{equation}
This yields the low temperature limit
\begin{equation}\label{LTB}
{F(\beta)\over L^2}={7\over 8}\times{\pi^2\over 720
d^3}-\left({1\over\pi\beta^3}+{1\over 2d\beta^2}\right) e^{-\pi\beta/2d}.
\end{equation}
In the low temperature limit the Helmholtz free energy for the original
Casimir's setup is given by, see {\it e. g.} Brown and Maclay
\cite{THERMALCASIMIR}
\begin{equation}\label{LTC}
{F(\beta)\over L^2}=-{\pi^2\over 720 d^3} -\frac{\zeta (3)}{2\pi\beta^3}
-\left({1\over\pi\beta^3}+{1\over d\beta^2}\right) e^{-\pi\beta/d}.
\end{equation}
Notice the absence of the factor proportional to $1/\beta^3$ in the case of
Boyer's setup. The reason for the absence of this factor will become clear
later on when we discuss the question of the temperature inversion symmetry
associated with the problem at hand.
The very high temperature limit is obtained by setting $\coth{\left({n\over
2\xi}\right)}\approx {2\xi\over n}$ and $\sinh{\left({n\over
2\xi}\right)}\approx {n\over 2\xi}$ and evaluating the sum. The result is:
\begin{equation}
f(\xi\gg 1)\approx {1\over 45}\pi^4\xi .
\end{equation}
This will lead to the Stefan-Boltzmann term corresponding to a slice of
vaccum of volume $L^2d$. More accurate  results at high temperature demand
that we transform the slowly convergent sum over $m$ in (\ref{DS}) into a
more rapidly convergent one. This can be accomplished with the help of
Poisson summation formula as we shall see next.
\section{The pressure} 
Let us go back to the scaled free energy $f(\xi)$ defined by the double sum
(\ref{DS}). Each of the double sums in equation (\ref{DS}) can be written in
the form
\begin{eqnarray}
& & \sum_{n,m=1}^\infty\left({a\over m^3}+{bn\over m^2}\right)e^{-nm/c} = 
\nonumber \\ \;\;\;\;\;\;\; & - & a\beta^2\sum_{n=1}^\infty 
\int_{n\kappa}^\infty d\omega\omega\ln\left({1-e^{-\beta\omega}}\right)\,
\end{eqnarray}
where $a$, $b$ and $c$ are constant satisfying the condition $a=bc$, and
$\kappa=1/\beta c$. The scaled free energy can be recast into the form
\begin{eqnarray}\label{HS}
f(\xi) & = & -\beta^2\sum_{n=1}^\infty \int_{n\kappa_1}^\infty
d\omega\;\omega\ln\left({1-e^{-\beta\omega}}\right) \nonumber \\
&+&\beta^2\sum_{n=1}^\infty
\int_{n\kappa_2}^\infty d\omega\;\omega\ln\left({1-e^{-\beta\omega}}\right)\,
\end{eqnarray}
where $\kappa_1=\pi/2d$ and $\kappa_2=\pi/d$. The first term (\ref{HS})
corresponds to the thermal correction for two infinite parallel perfectly
conducting ($\epsilon\to\infty$) plates separated by a distance $2d$. The
second one corresponds to the same setup but with the plates separated by a
distance $d$. In this way we can see that the setup we are considering here
is the difference between the two setups described above. If we apply this
reasoning to the zero temperature term, see (\ref{FREE}), we reproduce the
factor $7/8$ with the correct algebraic sign. The net pressure on the plates
is given by minus the derivative of the free energy per unit area with
respect to the distance $d$ between the plates and like the free energy it
splits into the zero temperature contribution and the thermal corrections,
that is
\begin{equation}
{\cal P}_{\mbox{\tiny net}}= {7\over 8}{\pi^2\over 240
d^4}+{1\over\pi^2\beta^4}{df(\xi)\over d\xi},
\end{equation}
The thermal contribution reads:
\begin{eqnarray}\label{TC}
{\cal P}_{\mbox{\tiny thermal}} &=& -{1\over\pi^2\beta^4\xi^3}\left[{1\over
4}\sum_{n=1}^\infty n^2\ln{\left(1-e^{-n/2\xi}\right)}\right. \nonumber \\
&-&\left. \sum_{n=1}^\infty n^2\ln{\left(1-e^{-n/\xi}\right)}\right].
\end{eqnarray}
Now we are ready to make use of one of the several versions of Poisson
summation formula \cite{Poisson}. The particular version suitable for our
purposes reads:
\begin{equation}\label{PoissonEven}
\sum_{n=1}^\infty G(n)=-{G(0)\over 2}+\sum_{l=-\infty}^\infty\int_0^\infty
dx\,G(x)\,\cos{(2\pi lx)},
\end{equation}
If we use (\ref{PoissonEven}) in (\ref{TC}) and add the result to the zero
temperature contribution we obtain
\begin{eqnarray}\label{NETPRESSURE}
{\cal P}_{\mbox{\tiny net}} & = & {\pi^2\over 45\beta^4}-{1\over
32\pi^4\beta^4}{\partial^2\over\partial\xi^2}{1\over\xi}
\sum_{m=1}^\infty{\coth{(4\pi^2 m \xi)}\over m^3} \nonumber \\
&+& {1\over 8\pi^4\beta^4}{\partial^2\over\partial\xi^2}{1\over\xi}
\sum_{m=1}^\infty{\coth{(2\pi^2 m \xi)}\over m^3}.
\end{eqnarray}
This result holds for all temperatures. Notice that the zero temperature
pressure is apparently missing in our final result. This happens because
upon the application of Poisson summation formula we obtain, besides the
Stefan-Boltzmann term and the two sums, a term with a negative sign which
exactly cancels out the repulsive zero temperature contribution. A similar
cancellation occurs also in the case of the Casimir effect for confined
massless fermions at finite temperature  \cite{GR} and in the high
temperature limit of the standard electromagnetic Casimir effect as shown,
for instance, in Plunien {\it et al} \cite{MosteTrunov}. Nevertheless, it is
a straightforward matter to show that if we take the zero temperature limit
of (\ref{NETPRESSURE}) we  recover the zero temperature term.

The high temperature limit is also easily obtained from (\ref{NETPRESSURE}).
Approximating conveniently the hyperbolic cotangents in the sums and
evaluating the second partial derivatives and keeping the leading correction
terms only we obtain:
\begin{eqnarray}
{\cal P}_{\mbox{\tiny net}}& \approx & {\pi^2\over \beta^4 45}+{3\zeta
(3)\over 16d^3\beta} \nonumber \\
&+&{1\over 2\pi d^3\beta}e^{-4\beta\pi d}
\left(1+{4\pi d\over\beta}+{8\pi^2 d^2\over\beta^2}\right).
\end{eqnarray}
 A simple integration of (\ref{NETPRESSURE}) yields another possible
representation for the Helmholtz free energy of this setup:
\begin{eqnarray}\label{HFE}
{F(\beta)\over L^2} & = & -{\pi^2d\over 45\beta^4}
+{1\over 32\pi^3\beta^3}{\partial\over\partial\xi}{1\over\xi}
\sum_{m=1}^\infty{\coth{(4\pi^2 m \xi)}\over m^3} \nonumber \\
&-& {1\over 8\pi^3\beta^3}{\partial\over\partial\xi}{1\over\xi}
\sum_{m=1}^\infty{\coth{(2\pi^2 m \xi)}\over m^3},
\end{eqnarray}
where the integration constant is determined by demanding that in the very
high temperature limit the only surviving term in (\ref{HFE}) must be the
Stefan-Boltzmann term. Notice that we can also determine this integration
constant analyzing the zero temperature limit of (\ref{HFE}). The high
temperature limit of (\ref{HFE}) is given by
\begin{eqnarray}\label{HTB}
{F(\beta)\over L^2} & \approx & -{\pi^2d\over 45\beta^4}+{3\over 32}{\zeta
(3)\over \pi d^2\beta}\\ \nonumber
& + & \left({1\over 4\pi d^2\beta}+{1\over d\beta^2}\right)e^{-4\pi d/\beta}
\end{eqnarray}
Apart from the all-important signs and numerical factors,
these results compare with those obtained in this limit for the attractive
case. From Brown and Maclay's results \cite{THERMALCASIMIR}, for example, we
can infer the following expression for the free energy per unit area in the
case of two infinite parallel perfectly conducting plates:
\begin{equation}
\frac{F(\beta)}{L^2}\approx -\frac{\pi^2 d}{45 \beta^4}-\frac{\zeta
(3)}{8\pi d^2\beta}-\left(\frac{1}{4\pi\beta
d^2}+\frac{1}{d\beta^2}\right)e^{-4\pi d/\beta}\,.
\end{equation}
In both cases in the very high temperature limit the dominant term is the
Stefan-Boltzmann term. Figures (\ref{FIGUREONE}) and (\ref{FIGURETWO}) show
the behavior of the scaled free energy function  $f(\xi):= [F(\beta)/L^2
d]\times d^4 $, where $F(\beta)/L^2$ is given by equation (\ref{HFE}) as a
function of the scaled temperature $\xi$. Calculations involved up to ten
terms in the summations required by (\ref{HFE}). We also show the scaled
free  energy corresponding to two perfectly conducting parallel plates as
given by equation (\ref{CONDPLATES}). Though exact, equation (\ref{HFE}) is
specially suited for the high temperature regime, convergence getting slower
and slower for small values of $\xi$. This can be seen if we examine in
greater detail the behavior of equation (\ref{HFE}) for very small values of
$\xi$. To obtain accurate results in this region it is necessary to include
many more terms in the required sums. Figure (\ref{FIGURETHREE}) shows the
behavior of the scaled pressure $p(\xi)$, which is given by equation
(\ref{NETPRESSURE}) multiplied by $d^4$, as a function of $\xi$. All curves
were plotted using MATHEMATICA version 2.2.1 \cite{MATHEMATICA}.  
\section{Temperature inversion symmetry}
Temperature inversion symmetry is a symmetry ocurring in the free energy
associated with the Casimir effect at finite temperature that depends on the
nature of the boundary conditions imposed on the quantum oscillations of the
massless field under study. In fact, as it was shown by Ravndal and
Tollefsen \cite{R&T89}, temperature inversion symmetry obtains for massless
bosonic fields and symmetric boundary conditions and for massless fermionic
fields and antisymmetric boundary conditions. One of the most remarkable
feature of the temperature inversion symmmetry is the possibilty of relating
the Stefan-Boltzmann term to the zero temperature Casimir effect.
Temperature inversion symmetry appears already in Brown and Maclay's work
\cite{THERMALCASIMIR} in their evaluation of the standard Casimir effect at
finite temperature. Brown and Maclay were able to write the scaled free
energy as a sum of three contributions: a zero temperature contribution {\it
i. e.}, the Casimir energy density at zero temperature, a Stefan-Boltzmann
contribution proportional the fourth power of the scaled temperature $\xi$,
and a non-trivial contribution. This non-trivial term is given
by\footnote{$\xi$ have the definition we have given herein. It differs from
$\xi$ in Brown and Maclay's paper by a factor $\pi$.}:
\begin{equation}
f_{\mbox{\tiny
non-trivial}}(\xi)=-\frac{1}{4\pi^2}\sum_{n=1}^\infty\,\sum_{m=1}^\infty
\frac{(2\pi\xi)^4}{[m^2+(2\pi n\xi)^2]^2}\,.
\end{equation}
This function has the following property:
\begin{equation}\label{TIS}
(2\pi\xi)^4 f\left(1/4\pi\xi\right)=f\left(\xi\right),
\end{equation}
which is the mathematical statement of the temperature inversion symmetry.
It turns out that the three contributions to the scaled free energy can be
combined
 and recasted into one piece as the double sum below:
\begin{equation}\label{CONDPLATES}
\tilde f(\xi)=-\frac{1}{16\pi^2}
\sum_{n=-\infty}^\infty\,\sum_{m=-\infty}^\infty
\frac{(2\pi\xi)^4}{[m^2+(2\pi n\xi)^2]^2}\,,
\end{equation}
where in the double sum we must exclude the term corresponding to $n=m=0$.
If we set $n=0$ and sum over $m$ with $m\neq 0$, we will obtain the
Stefan-Boltzmann term $-\pi^6\xi^4/45$. On the other hand, if we set $m=0$
and sum over $n$ with $n\neq 0$, we will obtain the Casimir term
corresponding to zero temperature $-\pi^2/720$.  As proved by Ravndal and
Tollefsen \cite{R&T89}, equation (\ref{CONDPLATES}) obeys the same symmetry
under temperature inversion as the one originally obtained by Brown and Maclay. 
It was also shown by Gundersen and Ravndal \cite{GR} that the scaled free
energy associated with massless fermions fields at finite temperature
submitted to MIT boundary conditions satisfy the relation given by equation
({\ref{TIS}) and therefore exhibts temperature inversion symmetry. Tadaki
and Takagi 
\cite{TT} have calculated Casimir free energies for a massless scalar field
obeying Dirichlet or Neumann boundary
conditions on both plates and found this symmetry. However, if the massless
scalar field must satisfy mixed boundary conditions, say, Dirichlet boundary
conditions on one plate and Neumann boundary conditions on the other, the
temperature inversion symmetry is lost. In the case of a massless scalar
field at finite temperature and periodic boundary conditions, it is possible
to show that the partition function, and consequently the free energy, can
be written in a closed form such that the temperature inversion symmetry
becomes explicit \cite{CWOT}.

In the repulsive case we are dealing with for which the boundary conditions
are not symmetric we should not expect to find this symmetry, nonetheless,
we will show that it is possible to extract a somewhat more complicated
version of the inversion temperature symmetry which will allow us to
establish a relationship between the high and the low temperature limits for
the repulsive case. We will also show that the underlying reason why it is
still possible to discuss the temperature inversion symmetry in our case is
the fact that Boyer's setup is equivalent to two original Casimir's setups,
a feature that we have already noted earlier (see equation \ref{HS}
).

Our starting point is equation (\ref{FR2}), which defines the dimensionless
function $f(\xi)$, and the identity:
\begin{equation}
\sum_{m=-\infty}^\infty{(-1)^m\over
\left(m^2+b^2\right)^2}={\pi^2\left[{1\over \pi b}+\coth{(\pi
b)}\right]\over 2b^2\sinh{(\pi b)}}\;,
\end{equation}
where here $b:=n/2\pi\xi$. Making use of the identity above we can rewrite
equation (\ref{FR2}) in the form
\begin{eqnarray}
f(\xi)&=&\frac{1}{8\pi^4\xi^3}
\sum_{m=-\infty}^\infty\,\sum_{n=1}^\infty
\left[\frac{(-1)^m(2\pi\xi)^4}{\left[n^2+(2\pi\xi
m)^2\right]^2}\right]\nonumber \\
&=&\frac{1}{16\pi^4\xi^3}\sum_{m=-\infty}^\infty\,
\sum_{n=-\infty}^\infty
\left[\frac{(-1)^m(2\pi\xi)^4}{\left[n^2+(2\pi\xi
m)^2\right]^2}\right]\nonumber \\
&-&\frac{1}{16\pi^4\xi^3}\sum_{m=-\infty}^\infty\frac{(-1)^m}{m^4}\,,
\end{eqnarray}
where in the double sum the term corresponding to $m=n=0$ and in the single
sum the term corresponding to $m=0$ must be both excluded. If we take the
expression above into equation (\ref{FREE}) we will be able to write the
free energy per unit area as
\begin{equation}\label{FREE2}
\frac{F}{L^2}=-\frac{1}{16\pi^2
d^3}\sum_{m=-\infty}^\infty\,\sum_{n=-\infty}^\infty\left[\frac{(-1)^m(2\pi\
xi)^4}{\left[n^2+(2\pi\xi m)^2\right]^2}\right]\,.
\end{equation}
Now we write down equation (\ref{FREE2}) as a sum over even terms plus a sum
over odd terms in $m$. If we do this we can separate the free energy per
unit area in two terms
\begin{equation}\label{FREESPLIT}
\frac{F}{L^2}=\frac{F_1}{L^2}-\frac{F_2}{L^2}\,, 
\end{equation}
where $F_1$ and $F_2$ are defined by
\begin{equation}
\frac{F_1}{L^2}:=-\frac{1}{128\pi^2
d^3}\sum_{m=-\infty}^\infty\,\sum_{n=-\infty}^\infty\left[\frac{(4\pi\xi)^4}
{\left[n^2+(4\pi\xi m)^2\right]^2}\right]\,,
\end{equation}
and
\begin{equation}
\frac{F_2}{L^2}:=-\frac{1}{16\pi^2
d^3}\sum_{m=-\infty}^\infty\,\sum_{n=-\infty}^\infty\left[\frac{(2\pi\xi)^4}
{\left[n^2+(2\pi\xi m)^2\right]^2}\right]\,.
\end{equation}
Now it is possible to verify directly that the free energies $F_1$ and $F_2$
satisfy the relations 
\begin{equation}\label{SYMMETRY1}
\left(4\pi\xi\right)^4F_1\left(\frac{1}{16\pi^2\xi}\right)=F_1\left(
\xi\right)\,,
\end{equation}
and
\begin{equation}\label{SYMMETRY2}
\left(2\pi\xi\right)^4F_2\left(\frac{1}{4\pi^2\xi}\right)=F_2\left(
\xi\right)\,.
\end{equation}
Equation (\ref{FREESPLIT}) can be construed as follows: First notice that
$F_1/L^2$ and $F_2/L^2$ can be respectively interpreted as the free energies
per unit area associated with two infinite parallel conducting plates
setups, one corresponding to a separation distance between the plates equal
to $2d$ and the other to a separation distance equal to $d$. In this way we
can say that the symmetries given by equations (\ref{SYMMETRY1}) and
(\ref{SYMMETRY2}) are induced by these correspondences, in accordance with
Ravndal and Tollefsen's \cite{R&T89} result, in the sense that both setups
require symmetric boundary conditions.

Observe also that by making use of the identity
\begin{equation}
\sum_{l=-\infty}^\infty\frac{1}{\left[
b^2+l^2\right]^2}=\frac{\pi}{2b^3}\coth{(\pi
b)}+\frac{\pi^2}{2b^2}\frac{1}{\sinh^2{(\pi b)}}\,,
\end{equation}
we can also write $F_1$ and $F_2$ as single sums
\begin{equation}
\frac{F_1}{L^2}=-\frac{\pi^2}{16
d^3}\sum_{n=-\infty}^\infty\left[\frac{4\xi^3}{n^3}\coth{\left(\frac{n}{4\xi}
\right)}
+\frac{\xi^2}{n^2}\frac{1}{\sinh^2{\left(\frac{n}{4\xi}\right)}}
\right]\,,
\end{equation}
and
\begin{equation}
\frac{F_2}{L^2}=-\frac{\pi^2}{16d^3}\sum_{n=-\infty}^\infty\left[\frac{4\xi^
3}{n^3}\coth{\left(\frac{n}{2\xi}
\right)}
+\frac{2\xi^2}{n^2}\frac{1}{\sinh^2{\left(\frac{n}{2\xi}\right)}}
\right]\,.
\end{equation}

As a first application of equations (\ref{FREESPLIT}), (\ref{SYMMETRY1}) and
(\ref{SYMMETRY2}) let us relate the Stefan-Boltzmann term, which is the
dominant term in the very high temperature limit, to the zero temperature
Casimir energy. In the very high temperature limit we can write for each setup
\begin{equation}
\frac{F_1(\infty)}{L^2}=-\frac{2}{45\beta^4}\pi^2d \,,
\end{equation}
and
\begin{equation}
\frac{F_2(\infty)}{L^2}=-\frac{1}{45\beta^4}\pi^2d\,.
\end{equation}
Making use of (\ref{SYMMETRY1}) and (\ref{SYMMETRY2}) we obtain
\begin{equation}
\frac{F_1(0)}{L^2}=-\frac{\pi^2}{4^3\times 90\, d^3}\,,
\end{equation}
and
\begin{equation}
\frac{F_2(0)}{L^2}=-\frac{\pi^2}{2^3\times 90\,d^3}\,.
\end{equation}
Hence, making use of (\ref{FREESPLIT}) we obtain
\begin{equation}
\frac{F_(0)}{L^2}=\frac{7}{8}\times\frac{\pi^2}{720\,d^3}\,.
\end{equation}
which is the Casimir energy at zero temperature associated with our original
setup. 

As a second application of our version of the temperature inversion symmetry
we now establish the relationship between the low and the high temperature
limits. In the high temperature limit the Helmholtz free energies $F_1$ and
$F_2$ corresponding to two setups each one of them formed by two infinite
parallel conducting plates kept at a distance $2d$ and $d$ apart,
respectively  read 
\begin{equation}\label{HTF1}
\frac{F_1}{L^2}\approx -\frac{2\pi^6\xi^4}{45 d^3}-\frac{\zeta
(3)\xi}{32d^3}-\left(\frac{\xi}{16d^3}+\frac{\pi^2\xi^2}{2d^3}\right)
e^{-8\pi^2\xi}\,.
\end{equation}
\begin{equation}\label{HTF2}
\frac{F_2}{L^2}\approx -\frac{\pi^6\xi^4}{45 d^3}-\frac{\zeta
(3)\xi}{8d^3}-\left(\frac{\xi}{4d^3}+\frac{\pi^2\xi^2}{d^3}\right)
e^{-4\pi^2\xi}\,,
\end{equation}
where we have made use of results obtained by Brown and Maclay for the
standard Casimir effect in the high temperature limit \cite{THERMALCASIMIR}.
Notice that if in accordance with (\ref{FREESPLIT}) we subtract (\ref{HTF2})
from (\ref{HTF1}) we will obtain the high temperature limit of the Helmholtz
free energy corresponding to Boyer's setup, equation (\ref{HTB}). Making use
of (\ref{SYMMETRY1}) and (\ref{SYMMETRY2}) we obtain
\begin{equation}
\frac{F_1}{L^2}=-\frac{\pi^2}{5 750d^3} -
\frac{\zeta(3)\pi^2\xi^3}{2d^3}-\left(\frac{\pi^2\xi^3}{d^3}+\frac{\pi^2\xi^
2}{2d^3}\right)
e^{-\frac{1}{2\xi}}\,,
\end{equation}
and
\begin{equation}
\frac{F_2}{L^2}=-\frac{\pi^2}{720 d^3} -
\frac{\zeta(3)\pi^2\xi^3}{2d^3}-\left(\frac{\pi^2\xi^3}{d^3}+\frac{\pi^2\xi^
2}{d^3}\right)
e^{-\frac{1}{\xi}}\,.
\end{equation}
Hence, upon making use of (\ref{FREESPLIT}) we obtain
\begin{equation}
\frac{F}{L^2}\approx \frac{7}{8}\frac{\pi^2}{720
d^3}-\left(\frac{\pi^2\xi^3}{d^3}+\frac{\pi^2\xi^2}{2d^3}\right)
e^{-\frac{1}{2\xi}}\,,
\end{equation}
which is the low temperature approximation to the Helmholtz free energy
corresponding to Boyer's setup, equation (\ref{LTB}).
\section{Conclusions}
In this paper we have shown how neatly the generalized zeta function
regularization method applies to the repulsive electromagnetic Casimir
effect at finite temperature for the simple geometry of a pair of infinite
parallel plates, each one of them endowed with a certain special physical
property, namely, perfect electric conduction and infinite magnetic
permeability. Advantage was taken from the fact that for this simple
geometry, the electromagnetic field can be simulated by a massless uncharged
scalar field. As a follow-up to the application of this method we have
obtained expressions for the Helmholtz free energy and the force per unit
area acting on any one of the two plates which comprise this peculiar
system. We have obtained the low and high temperature limits of those two
quantities.  We have also shown that though Boyer's plates demand the
imposition of non-symmetric boundary conditions on the scalar massless field
that simulates the electromagnetic field, it is still possible to take
advantage of the temperature inversion symmetry, a symmetry which in
principle does not hold for the case we have studied here, and relate the
high and the low temperature limits, particularly the Stefan-Boltzmann free
energy and the Casimir energy at zero temperature.  

It is also worth noticing that our high temperature limit result is
compatible with ideas of dimensional reduction that occurs for $\beta$ $\to
0$. Remark that in this limit, if we omit the Stefan-Boltzmann term,  the
dominant term in the pressure is proportional to $1/d^3$, a result
reminiscent of $2+1$ dimensions, in contrast to the $1/d^4$ term, typical of
$3+1$ dimensions. 

The interesting similarities and differences of the Casimir effect
associated with a massless bosonic field, which arise when we compare the
consequences of imposing Dirichlet boundary conditions in one case and mixed
ones in the other, indicate that a similar investigation in other theories,
such as the massive scalar field at zero as well as at finite temperature
might be rewarding. This investigation is being carried out and results will
be published elsewhere.

\acknowledgments
\rm The authors are indebt to our colleagues Carlos Farina and Ashok Das for
carefully reading the manuscript and enlightning discussions. We wish to
acknowledge also our colleague V. Soares for helping us with the figures.
One of us (A. Ten\'orio) wishes to acknowledge the financial support of
CNPq, the Brazilian research agency.
\clearpage
\begin{figure}
\caption{The scaled free energy per unit area as a function of the scaled
temperature for two perfectly conducting parallel plates and for Boyer's
unusual setup. The scaled free energy is given by $f(\xi):=
d^4\times[F(\beta)/L^2 d]$, where $F(\beta)/L^2$ is given by equation
(\ref{HFE}). The scaled temperature $\xi$ is defined by $\xi:=d/\pi\beta $.
The Casimir energy at zero temperature for the parallel conducting plates
setup and Boyer's setup are represented respectively by the straight lines
parallel to the scaled temperature axis intercepting the vertical axis at
$f(0) =-\pi^2/720 \approx -0.014$ and $f(0)=(7/8)\times \pi^2/720 \approx
0.012$. The scaled free energy curve for Boyer's setup is represented by
longer dashes and tends to $(7/8)\pi^2/720\approx 0.012$ when $\xi\to 0$.}
\label{FIGUREONE}
\end{figure}
\begin{figure}
\caption{The scaled free energy per unit area as a function of the scaled
temperature for two perfectly conducting parallel plates and Boyer's setup
for higher values of the scaled temperature. The scaled free energy for
Boyer's setup is the curve represented by longer dashes. In the high $\xi$
limit both curves tend to the Stefan-Boltzmann curve, $-\pi^6\xi^4/45$.}
\label{FIGURETWO}
\end{figure}
\begin{figure}
\caption{The scaled pressure for Boyer's setup. The vertical axis represents
the dimensionless function $p(\xi):= d^4\times\cal{P}_{\mbox{\tiny
net}}(\beta)$, where $\cal{P}_{\mbox{\tiny net}}$ is given by equation
(\ref{NETPRESSURE}). The straight line parallel to the $\xi$-axis intercepts
the scaled pressure axis at $p(0)=(7/8)\pi^2/240 \approx 0.036$.}
\label{FIGURETHREE}
\end{figure}

\begin{thebibliography}{*}
%
\bibitem[*]{email}  \textrm{Electronic-mail address: tort@if.ufrj.br }
%
\bibitem{Ca} H. B. G. Casimir, Proc. K. Ned. Akad. Wet. {\bf 51}, 793 (1948).
%
\bibitem{MosteTrunov} V. M. Mostepanenko and N. N. Trunov, Sov. Phys. Usp.
{\bf 31}, 965 (1988); V. M. Mostepanenko and N. N. Trunov, {\it The Casimir
Effect and its Applications}, Oxford: Clarendon (1997); G. Plunien, B.
M\"uller and W. Greiner, Phys. Rep. {\bf 134}, 664 (1987).
%
\bibitem{Lamo} S. K. Lamoreaux, Phys. Rev. Lett. {\bf 79}, 5 (1997); U.
Mohideen and A. Roy, Phys. Rev. Lett. {\bf 81}, 4549 (1998)
%
\bibitem{Actor} J. R. Ruggiero, A. H. Zimerman and A. Villani, Rev. Bras.
F\'{\i}s. {\bf 7}, 663 (1977);
 A. A. Actor,  Phys. Rev. {\bf D50}, 6560 (1994); Phys. Rev. {\bf D52}, 3581
(1995).
%
\bibitem{ZETA} A. Salam and J. Strathdee, Nuc. Phys B {\bf 90}, 203 (1975);
J. S. Dowker and R. Critchley, Phys. Rev. D {\bf 13}, 3224 (1976); S. W.
Hawking, Commun. Math. Phys. {\bf 55}, 133 (1977); G. W. Gibbons, Phys.
Lett. A {\bf 60}, 385 (1977);  See also: E. Elizalde, S. D. Odintsov, A.
Romeo, A. A. Bytsenko and S. Zerbini {\it Zeta Regularization Techniques
with Applications}, World Scientific, Singapore (1994).
%
\bibitem{Boyer} T. H. Boyer, Phys. Rev. {\bf A9}, 2078 (1974).
%
\bibitem{CP} M. V. Cougo-Pinto, C. Farina and A. Ten\'orio: {\it Zeta
Function Method for the Repulsive Casimir Effect}. To appear in {\it
Brazilian Journal of Physics}.
%
\bibitem{E&R91} E. Elizalde and A. Romeo, Am. J. Phys. {\bf 59}, 711 (1991). 
%
\bibitem{Hushwater} V. Hushwater, Am. J. Phys. {\bf 65}, 1627 (1997).
%
\bibitem{Scharnhorst} K. Scharnhorst, Phys. Lett {\bf B236}, 354 (1990).
{\it See also}: G. Barton, {\it Phys. Lett.} {\bf B237}, 559 (1990).
%
\bibitem{CPFST} M. V. Cougo-Pinto, C. Farina, F. C. Santos and A. C. Tort,
{\it The speed of light in confined QED vacuum: faster or slower than $c$?}.
To appear in Phys. Lett. {\bf B};  M. V. Cougo-Pinto, C. Farina, F. C.
Santos and A. C. Tort, {\it QED vacuum between an unusal pair of plates},
hep-th/9811062.
%
\bibitem{THERMALCASIMIR} M. Fierz, Helv. Phys. Acta {\bf 33} 855 (1960); F.
Sauer, Dissertation, Universit\"{a}t G\"{o}ttingen (1962); J. Mehra, Physica
{\bf 37}, 145 (1967); L. S. Brown and G. J. Maclay Phys. Rev. {\bf 184} 1272
(1969); {\it See also} J. Schwinger, K. A. Milton and L. L. DeRaad Jr., Ann.
Phys. {\bf 115}, 1 (1978).  
%
\bibitem{Kapusta} A. Das, Finite Temperature Field Theory, World Scientific,
Singapore (1997).
%
\bibitem{Grad} I. S. Gradshteyn and I. M. Ryzhik, Tables of Integrals,
Series and Products, 5th Edition, Academic Press, New York (1994).
%
\bibitem{Epstein} P. Epstein, Math. Ann. {\bf 56}, 615 (1903);  Math. Ann.
{\bf 63}, 205 (1907).
%
\bibitem{Kirsten} E. Elizalde and A. Romeo, J. Math. Phys. {\bf 30} 1133,
(1989); {\it erratum} {\bf 31}, 771 (1990); K. Kirsten, J. Math. Phys. {\bf
35}, 459 (1994).
%
\bibitem{Poisson} S. D. Poisson, Journal de L'\'Ecole Polytechnique {\bf
XII}, Cahier XIX, 420 (1823).
%
\bibitem{GR} S. A. Gundersen and F. Ravndal, Ann. of Phys. {\bf 182}, 90
(1988). 
%
\bibitem{MATHEMATICA} S. Wolfram and G. Beck, {\it Mathematica; The Student
Book}, Addison-Wesley Pub. Co. Inc. 
Reading (1994); R. L. Zimmerman and F. I. Olness, {\it Mathematica for
Physics} Addison-Wesley Pub. Co. Inc. 
Reading (1995).
%
\bibitem{R&T89} F. Ravndal and D Tollefsen, Phys. Rev. D {\bf 40}, 4191 (1989).
%
\bibitem{CWOT}  C. Wotzasek, J. Phys. A {\bf 23}, 1627 (1990).
%
\bibitem{TT} S. Tadaki and S. Takagi, Progr. Theor. Phys. {\bf 75} 262 (1986).
%
\end{thebibliography}
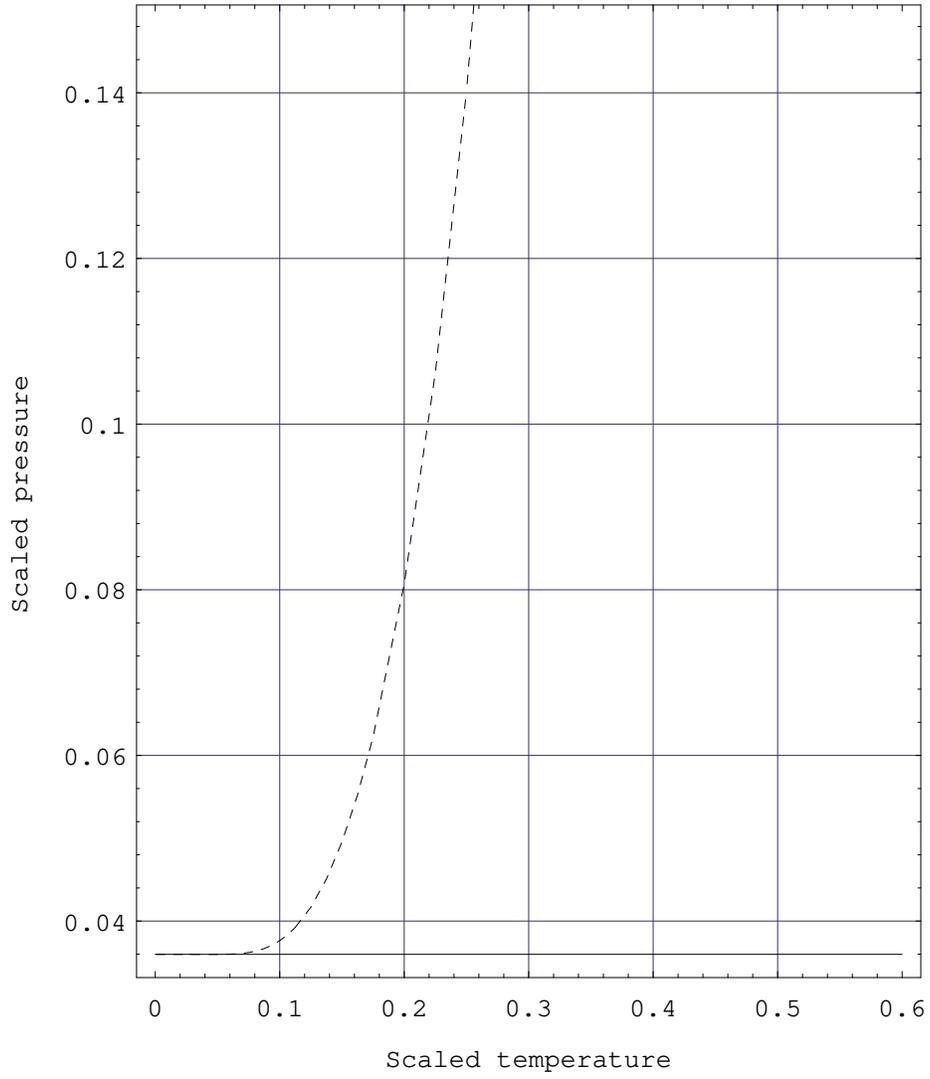
\end{document}